\documentclass[12pt]{article}

\usepackage{lineno}
\usepackage[margin=1in]{geometry}
\usepackage{amssymb,amsmath,latexsym}                        
\usepackage{graphicx}
\usepackage{cite}
\usepackage{xcolor}
\usepackage{float}

\usepackage{amsthm}
\usepackage{multicol}
\usepackage{authblk}
\usepackage{subfigure}

\usepackage[labelfont=bf]{caption}

\usepackage[colorlinks=true,linkcolor=blue, citecolor=blue, urlcolor=blue]{hyperref}

\usepackage{adjustbox}
\usepackage{bbm}
\usepackage{mathtools}

\date{\today} 

\begin{document}

\title{The Optimal Use of Silicon Pixel Charge \\ Information for Particle Identification }
\author[1]{Harley Patton}
\author[2]{Benjamin Nachman}
\affil[1]{\normalsize\it Computer Science Division, University of California, Berkeley}
\affil[2]{\normalsize\it Physics Division, Lawrence Berkeley National Laboratory}

\maketitle

\begin{abstract}
Particle identification using the energy loss in silicon detectors is a powerful technique for probing the Standard Model (SM) as well as searching for new particles beyond the SM.  Traditionally, such techniques use the truncated mean of the energy loss on multiple layers, in order to mitigate heavy tails in the charge fluctuation distribution.  We show that the optimal scheme using the charge in multiple layers significantly outperforms the truncated mean.  Truncation itself does not significantly degrade performance and the optimal classifier is well-approximated by a linear combination of the truncated mean and truncated variance.
\end{abstract}

%The distribution of the clusters should also be useful - think about this also for the slow proton data?

%-------------------------------------------------------------------------------
\section{Introduction}
\label{sec:intro}
%-------------------------------------------------------------------------------

Charged hadron identification plays a key role in many collider-based particle and nuclear physics analyses.   For example, the ATLAS~\cite{Aad:2008zzm,ATLAS-CONF-2011-016}, CMS~\cite{Chatrchyan:2008aa,Khachatryan:2010pw}, and ALICE~\cite{Aamodt:2008zz,Abelev:2014ffa} experiments\footnote{The LHCb experiment~\cite{Alves:2008zz} measures ionization in silicon from charged-particle tracks passing through their VELO detector~\cite{LHCbVELOGroup:2014uea}.  This information is used for improving the position resolution~\cite{Parkes:1074928} but only exploratory work exists for using it for particle identification~\cite{McGregor:2011fna}.} at the Large Hadron Collider (LHC) can use the ionization energy loss (dE/dx) in silicon\footnote{Ionization energy loss can be measured in other systems as well, for instance demonstrated in straw tubes in the ATLAS experiment's Transition Radiation Tracker (TRT)~\cite{1748-0221-12-05-P05002}.  The ideas discussed in this paper apply also to these other applications, but for focus the discussion here is restricted to silicon.} from charged-particle trajectories to classify particles at low momentum ($\beta\gamma\lesssim 1$).   This particle identification capability has been used to study properties of hadronization~\cite{Chatrchyan:2012qb,Chatrchyan:2013eya}, including the effects of Bose-Einstein correlations~\cite{cmsbecs} as well as to search for massive long lived unstable particles (LLPs), highly ionizing particles (HIPs), and heavy stable charged particles (HSCPs)~\cite{Aaboud:2016dgf,Khachatryan:2016sfv,ATLAS:2014fka,Chatrchyan:2013oca,Aad:2011yf,Aad:2012pra,Aad:2013pqd,Chatrchyan:2012sp,Khachatryan:2011ts} that are predicted in many theories of physics beyond the Standard Model.  Particles with low $\beta\gamma$ or high electric charge can deposit significantly more energy than minimum ionizing particles (MIPs), with the average charge\footnote{There is a significant difference between the mean and mode; the mode contains more information about the particle identity.  For a comprehensive review about this topic and corrections to Eq.~\ref{eq:bethe}, see e.g. Ref.~\cite{Bichsel:2006cs,Patrignani:2016xqp}.} following the Bethe equation:

\begin{align}
\label{eq:bethe}
%\left\langle -\frac{\text{dE}}{\text{dx}}\right\rangle& \approx \rho Kz^2\frac{Z}{A}\frac{1}{\beta^2}\left[\log\left(\frac{2m_e\beta^2\gamma^2}{I}\right)-\beta^2\right]\\
%& \approx \frac{0.35}{\beta^2}\left[8.7+2\log\left(\beta\gamma\right)-\beta^2\right]\text{ MeV / cm}, \hspace{5mm}\text{(silicon)}
\left\langle -\frac{\text{dE}}{\text{dx}}\right\rangle& \approx \frac{0.35}{\beta^2}\left[8.7+2\log\left(\beta\gamma\right)-\beta^2\right]\text{ MeV / cm}.
\end{align}

\noindent Quoted values of dE/dx often divide by the density $\rho\approx 2.3 \text{ g}/\text{cm}^3$ to report a value in MeV cm$^{2}$ / g.  For example, with a momentum of 500 MeV, the relative average dE/dx of protons, kaons, and pions are 3, 1.5, 1, respectively. 

A key challenge with dE/dx-based particle-identification is that the energy loss probability distribution has significant and asymmetric fluctuations (`straggling').   As a result of the skewed energy loss fluctuations (approximately Landau-distributed), the mean energy loss is much higher than the most probable energy loss.  In addition to causing primary energy loss, ionized electrons can also have sufficient energy to cause further energy loss by ionization or excitation ($\delta$-rays or knock-out electrons).  These $\delta$-rays are slow moving and highly ionizing, especially near their Bragg peak.  For a complete review of energy loss fluctuations in silicon, see e.g. Ref.~\cite{Bichsel:2006cs,Patrignani:2016xqp}.  To overcome the challenge of a significantly different energy loss straggling mean and mode, dE/dx-based methods combine information from multiple silicon layers and traditionally have used a truncated mean to approximate the mode of the charge distribution~\cite{Aad:2008zzm,ATLAS-CONF-2011-016,Chatrchyan:2008aa,Khachatryan:2010pw,Aamodt:2008zz,Abelev:2014ffa,Chatrchyan:2012qb,Chatrchyan:2013eya,cmsbecs,Aaboud:2016dgf,Khachatryan:2016sfv,ATLAS:2014fka,Chatrchyan:2013oca,Aad:2011yf,Aad:2012pra,Aad:2013pqd,Chatrchyan:2012sp,Khachatryan:2011ts}.   Truncating and only using the average both remove information that may be useful for particle identification.  The purpose of this paper is to study how much information, if any, is lost by these two standard data reduction schemes.  

%can measure the trajectory of these particles with tracking detectors that are composed of silicon pixels at their core\footnote{The ALICE and LHCb experiments also measure tracks, but mostly rely on other detectors for particle identification.  ALICE does use the energy loss in silicon to cover some holes in phase space not well-covered by their other tracking detectors~\cite{Abelev:2014ffa}.  Actually, the use the truncated mean ... maybe we should put that up in the main body.}.  The ionization energy loss pattern, measured with the time-over-threshold (ToT) technique~\cite{603658}, depends on the hadron momentum as well as its type.  When the momentum is below about 1 GeV, it is possible to use the energy loss to differentiate pions, protons, and kaons.  This has been used in a variety of measurements aimed at studying the properties of hadronization~\cite{Chatrchyan:2012qb,Chatrchyan:2013eya}, including the effects of Bose-Einstein correlations~\cite{cmsbecs}.    Next is long-lived particle searches.

%In addition, there is an entire class of searches at the LHC for physics beyond the Standard Model that exploit the charge measurement of pixel detectors.  This includes searches for massive long lived unstable particles (LLPs), highly ionizing particles (HIPs), and heavy stable charged particles (HSCPs)~\cite{Aaboud:2016dgf,Khachatryan:2016sfv,ATLAS:2014fka,Chatrchyan:2013oca,Aad:2011yf,Aad:2012pra,Aad:2013pqd,Chatrchyan:2012sp,Khachatryan:2011ts}.

%Should have a paragraph here about what has been done in the past and what we will do differently in this paper.

This paper is organized as follows.  Section~\ref{sec:simulation} introduces the simulation framework, and existing methods based on the truncated mean are illustrated in Sec.~\ref{sec:existmeth}. The improvements from using all of the charge information are demonstrated in Sec.~\ref{sec:optclass}.  The paper ends with some conclusions in Sec.~\ref{sec:concl}.
 
%-------------------------------------------------------------------------------
\section{Simulation}
\label{sec:simulation}
%-------------------------------------------------------------------------------

\begin{figure}[h]%
	\centering
	\includegraphics[width=0.6\textwidth]{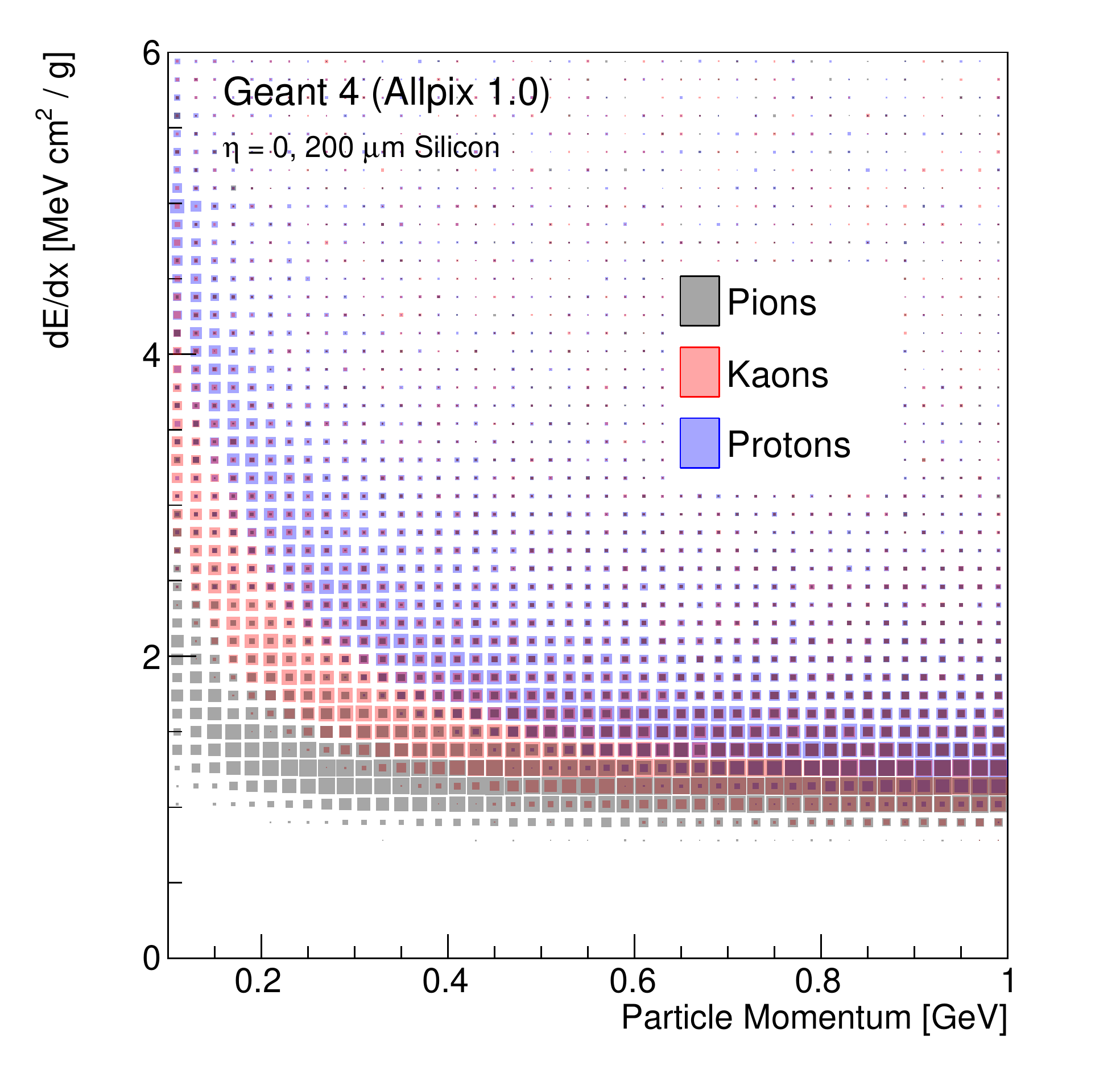}%
	\caption{The distribution of the pixel cluster dE/dx as a function of the particle momentum for pions, kaons, and protons.}
	\label{fig:2d}
\end{figure}

For concreteness, a detector setup similar to the ATLAS pixel detector is used for illustration.  There are four pixel layers, each $200$ $\mu$m thick with a pitch of $50\times 250$ $\mu$m$^2$.  Particles of a given momentum are incident perpendicular to the pixel surface\footnote{Charge fluctuations depend on the sensor thickness.  Since the path length in silicon is the same for all particles in this study, this effect is removed.  Future work could study the improvements to dE/dx-based tagging with variable path lengths (incidence angles) and adding thickness as a discriminating feature.}.   Detector geometry and particle propagation are simulated using Allpix~\cite{benoit:20xx}, built on the Geant4 package~\cite{Agostinelli:2002hh}.   The setup is identical to the one described in Ref.~\cite{Chen:2017yzr} and is summarized here for completeness.  Charge depositions and fluctuations are provided by Geant4 using the \textsc{emstandard\_opt0} model\footnote{This is not an accurate model for thin sensors, but $200$ $\mu$m are sufficiently thick that the total deposited charge is well-modeled~\cite{Wang:2017ygj}.}.  The ionization energy is converted into electron-hole pairs and electrons are transported to the collecting electrode, including drift and diffusion.  The diffusion length scales with the square-root of the drift time~\cite{ANDP:ANDP18551700105} and the diffusion constant is modeled according to the Einstein relation~\cite{einsteinrelation1,einsteinrelation2,einsteinrelation3}.   Electron and hole mobilities are parameterized using the common Canali-modified Caughey and Thomas velocity saturation model~\cite{Caughey1967,Canali1975}.  The temperature is set to $273$ K.  In addition to diffusion, charges are deflected in a 2 T magnetic field that is perpendicular to the sensor depth.  The angle of deflection is the Lorentz angle, given by $\tan\theta =r \mu B$.  Deposited energy is digitized using a the time-over-threshold (ToT) method, with a linear conversion.  The analog threshold is 3000 electrons, there are 8 bits of ToT, and a minimum ionizing particle (MIP) at perpendicular incidence corresponds to a ToT of 128 (half of the available range).  The ToT is then converted to dE/dx by assuming 80 e/h pairs per micron for a MIP and 3.6 eV per e/h pair.  Figure~\ref{fig:2d} shows the pixel cluster dE/dx distribution as a function of particle momentum for pions, kaons, and protons.  Since protons are more massive than kaons which are more massive than pions, the proton dE/dx is shifted to higher values than kaons which is shifted to higher values than pions.  For illustration, the next sections will focus on the kaon-versus-pion classification task; protons will also be discussed at the end of Sec.~\ref{sec:optclass}.

\clearpage

%-------------------------------------------------------------------------------
\section{Existing Methods}
\label{sec:existmeth}
%-------------------------------------------------------------------------------

\begin{figure}[h!]%
	\centering
	\subfigure{%
		\includegraphics[width=0.45\textwidth]{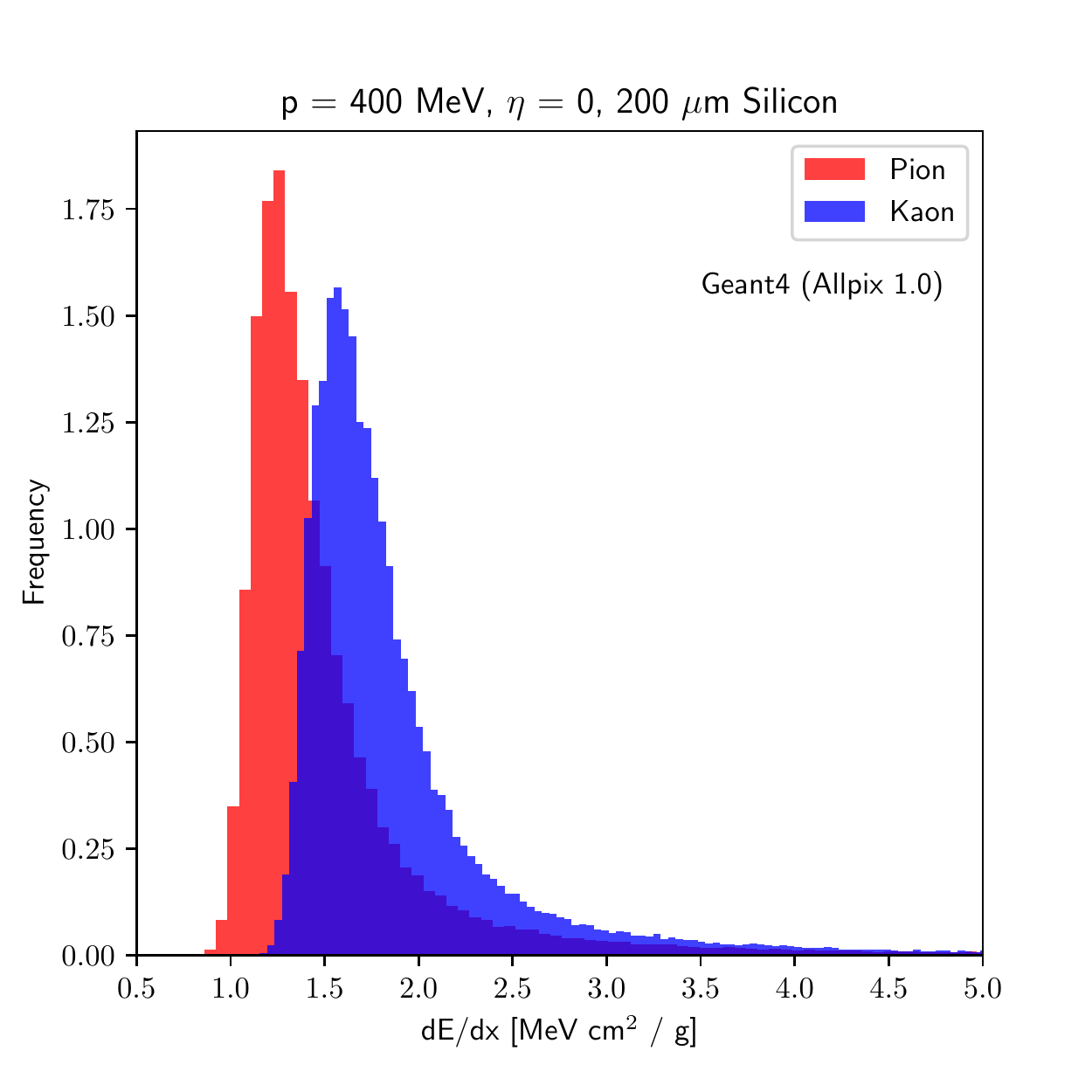}}%
	\qquad
	\subfigure{%
		\includegraphics[width=0.45\textwidth]{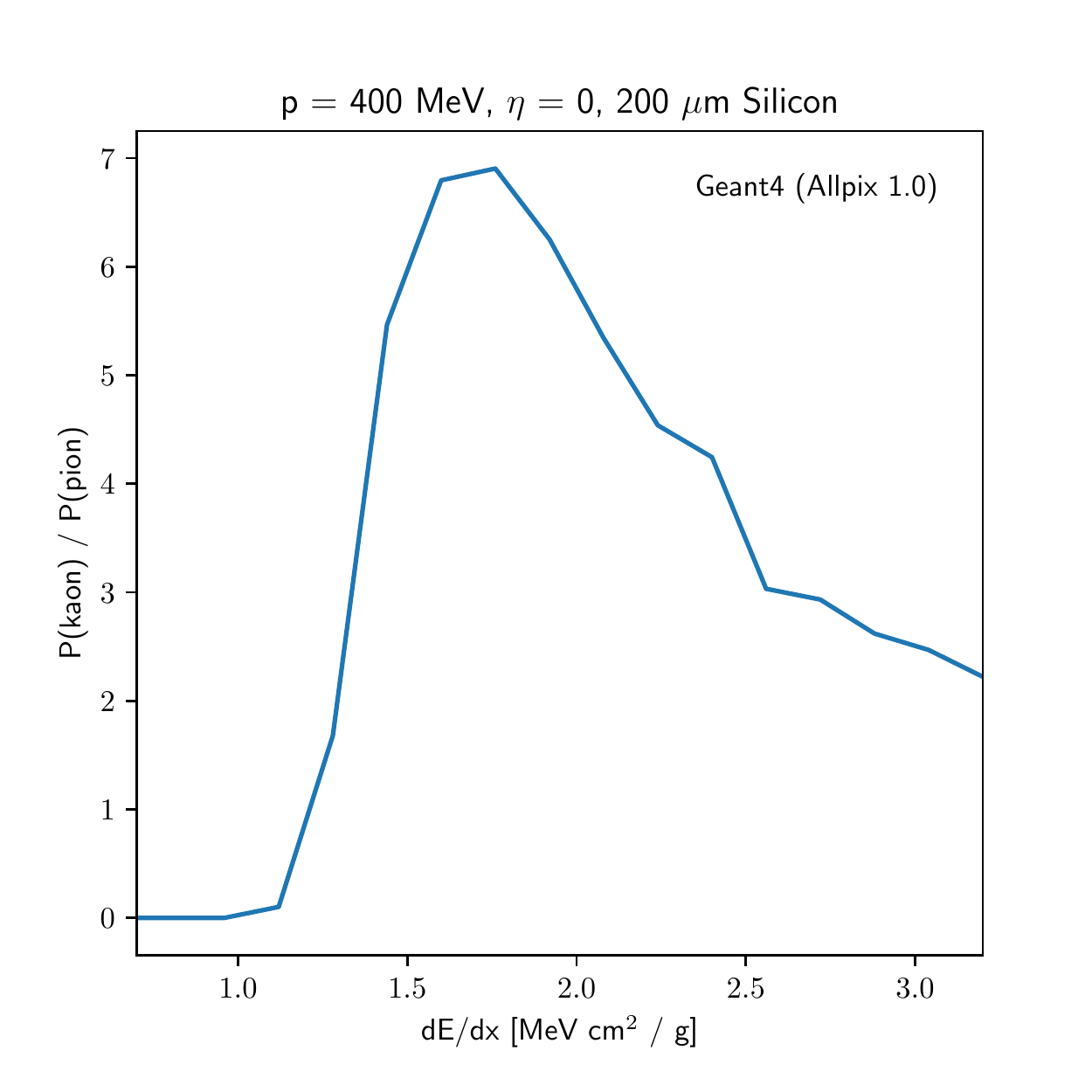}}%
	\caption{The distribution of energy loss for pion and kaon tracks (left) at a momentum of 400 MeV follows the Landau distribution, with a likelihood ratio (right) which levels off at higher energy loss values.}
	\label{fig:why_we_truncate}
\end{figure}

For $n$ dE/dx measurements $\{Q_i\}_{i=1..n}$, the truncated mean $\mu_t$ is simply the average over all the values, excluding $\text{max}_i Q_i$.  If more than two $Q_i$ are the maximum, then multiple $Q_i$ are discarded.  This is the most standard approach to dE/dx-based charge identification.  The reasoning behind this choice is that the energy loss in each detector layer is nearly Landau-distributed, with a very heavy tail that carries little to no discriminatory power.   The left plot of Fig.~\ref{fig:why_we_truncate} shows the dE/dx distribution for pions and kaons with $p=400$ MeV.  As expected, the kaon distribution is shifted to the right of the pion distribution because $\beta\gamma$ is lower for the kaons due to their higher mass.  The right plot of Fig.~\ref{fig:why_we_truncate} shows the likelihood ratio of kaons to pions.  Where the likelihood ratio is not unity, the distribution has significant discrimination power.  Beyond dE/dx $\sim 3$, the ratio is rather flat and close to $1$.  This is largely dominated by $\delta$-rays which have nearly the same spectrum for different particle species of the same momentum.  

\begin{figure}[h!]%
	\centering
		\subfigure{%
			\includegraphics[width=0.45\textwidth]{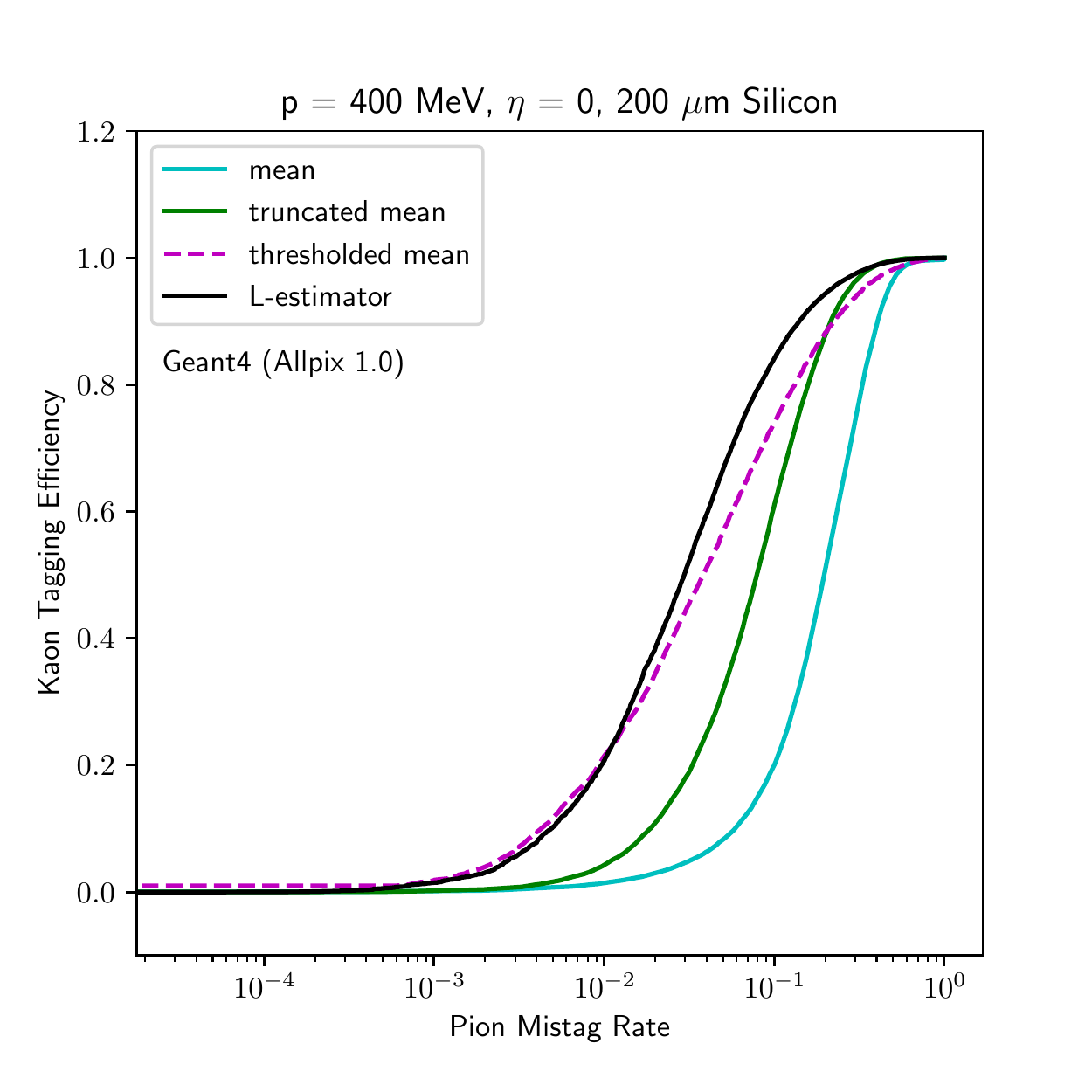}}%
		\qquad
		\subfigure{%
			\includegraphics[width=0.45\textwidth]{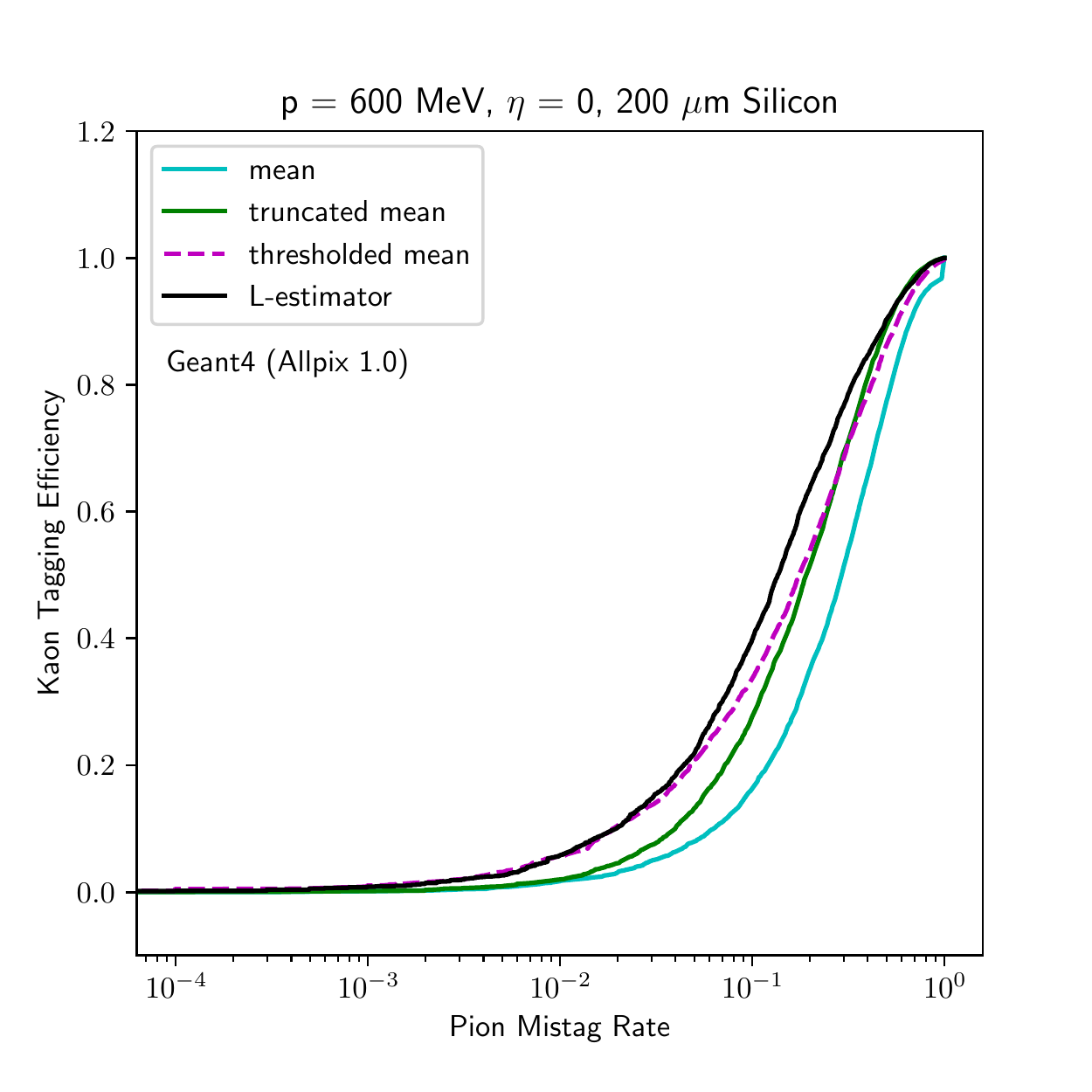}}%
	\caption{Truncation, thresholding ($\theta = 1.8\, \text{MeV}\, \text{g}^{-1} \text{cm}^2$), and the optimal L-estimator all provide a significant boost in mean-based classifier performance, shown here at momenta of both 400 MeV (left) and 600 MeV (right).}
	\label{fig:truncate}
\end{figure}

Figure~\ref{fig:truncate} shows the classification performance for various types of (generalized) truncated mean.   A general L-estimator~\cite{Talman:1978tc,Regeler:981340} is given by

\begin{align}
\mathcal{L}_\text{L-est}(\vec{x}) = \sum_i w_i x_{(i)},
\end{align}

\noindent where $\sum w_i=1$ and $x_{(i)}$ are the $i^\text{th}$ order statistics: $x_{(1)} < x_{(2)}<\cdots < x_{(n)}$ for $n$ measurements $x_i$.  The truncated mean is a special case given by $w_n=0$ and $w_i=1/(n-1)$ for $i<n$.  The optimal L-estimator was empirically determined by training a linear classifier on the order statistics using scikit-learn~\cite{scikit-learn}, and corresponds to $w\approx (0.778,0.231,-0.005,-0.004)$.  Its performance is quantified in Fig.~\ref{fig:truncate} with a receiver operating characteristic (ROC) curve where the probability to correctly tag a kaon is traded off with the probability to incorrectly label a pion as a kaon.  The curve traces out various thresholds on the mean dE/dx.  The usual truncated mean significantly out-performs the regular mean over all layers for a pion mistag rate above about 1\%.  However, Fig.~\ref{fig:why_we_truncate} suggests that maybe truncation is actually too coarse; dE/dx is only unhelpful for values above about $2$ MeV cm$^2$/g.  Therefore, a second truncation scheme is used whereby all values of dE/dx are used so long as they are less than some threshold $\theta$.  This scheme out-performs the truncated mean between 0.1\% and about 10\% pion mistag rates.  Figure~\ref{fig:truncate} uses $\theta=1.8$ MeV cm$^2$/g.  A lower threshold value will result in a more symmetric energy loss distribution (unless it is too low), but may throw away potentially valuable information about the spread of the values.  This new scheme does introduce a new hyper-parameter; the next section considers taking this to the extreme by trying to approximate the optimal classifier which has many hyper-parameters that can be learned from the simulation (or data).  The L-estimator in Fig.~\ref{fig:truncate} already matches or out-performs the other classifiers, which suggests that there is more useful information than just the mean.  In particular, the optimal L-estimator is approximately $\frac{1}{2}(x_{(1)}+x_{(2)}) - \frac{1}{4}(x_{(2)}-x_{(1)})$, which resembles difference between a truncated mean and a measure of `truncated spread', suggesting that information on both quantities may be necessary for optimal classification - discussed more in Sec.~\ref{sec:optclass}.

\clearpage

\clearpage

%-------------------------------------------------------------------------------
\section{Optimal Classification}
\label{sec:optclass}
%-------------------------------------------------------------------------------

% By the Neyman-Pearson lemma~\cite{nplemma}, an optimal classifier is the likelihood ratio: $h_\text{optimal}(\vec{x}) = p_S(\vec{x})/p_B(\vec{x})$.  Therefore, the goal of classification is to learn $h_\text{optimal}$ or any classifier that is monotonically related to it.  

%%%%%

In the context of particle identification, each track corresponds to a $n$-dimensional sample point $\vec{x} \in \mathbb{R}^n$ with a feature representing the energy loss readout ($Q_i$) for each of the $n$ detector layers:

\vspace{-5mm}

\begin{align}
\vec{x} = \begin{pmatrix}
Q_1 & Q_2 & \dots & Q_n
	\end{pmatrix}.
\end{align}

\noindent The goal of particle identification is to distinguish one process (called $S$) from another process (called $B$).  For illustration, $S$ will correspond to kaons and $B$ will represent pions.   A classifier is a real-valued function $h$ that is optimized to give different values when presented examples of $\vec{x}$ from $S$ and $B$.  The classifier $h$ is optimal if the corresponding ROC curve is no worse than any other classifier, i.e. for any other classifier $h'$, the probability to misclassify $B$ as $S$ at a fixed signal efficiency for $h'$ is no lower than for $h$ at all possible signal efficiencies.  By the Neyman-Pearson lemma~\cite{nplemma}, an optimal classifier exists and is given by thresholding the likelihood ratio\footnote{Note that this is not the same as the maximum likelihood estimator, $\mathcal{L}_\text{ML}(\vec{x})=\max_jp_j(\vec{x})$.  The classifier $\mathcal{L}_\text{ML}$ has been studied in the past (see e.g. Ref.~\cite{Regeler:981340}) but is provably no better than the optimal classifier and in previous studies was found to be worse than $\mathcal{L}_\text{L-est}$ and thus is not considered further.} $\mathcal{L}(\vec{x})_\text{opt}=p_S(\vec{x})/p_B(\vec{x})$, where $p_S(\vec{x})$ is the probability density for $\vec{x}$ for $S$ and similarly for $B$.  For MIPs going through a pixel detector, $p(\vec{x})$ is well-approximated as a product over the probabilities for $Q_i$ in each layer separately.  Even with this decomposition, it can be difficult to visualize and validate the optimal classifier in $n$ dimensions and so there is great utility in having a classifier in a reduced feature space where the optimal classifier is specified by a linear decision boundary.  Our goal is to both identify the optimal classifier and attempt to find a low-dimensional approximation where a linear threshold is close to optimal.

%A classifier is a statistical model that can identify to which of a set a categories a sample point belongs, and is said to be Bayes optimal when its probabilistic error over all sample points is at a minimum. The predictions returned by optimal classifiers acting in the original feature space of the data can often be hard to reason about, due to both the high dimensionality of the sample points and the nonlinearity of the resulting decision boundaries. Therefore, it is of great use to find a transformation that can map the sample points to new feature space that is both of lower dimensionality and contains a linear decision boundary.

\begin{figure}[h]%
	\centering
		\subfigure{%
			\includegraphics[width=0.45\textwidth]{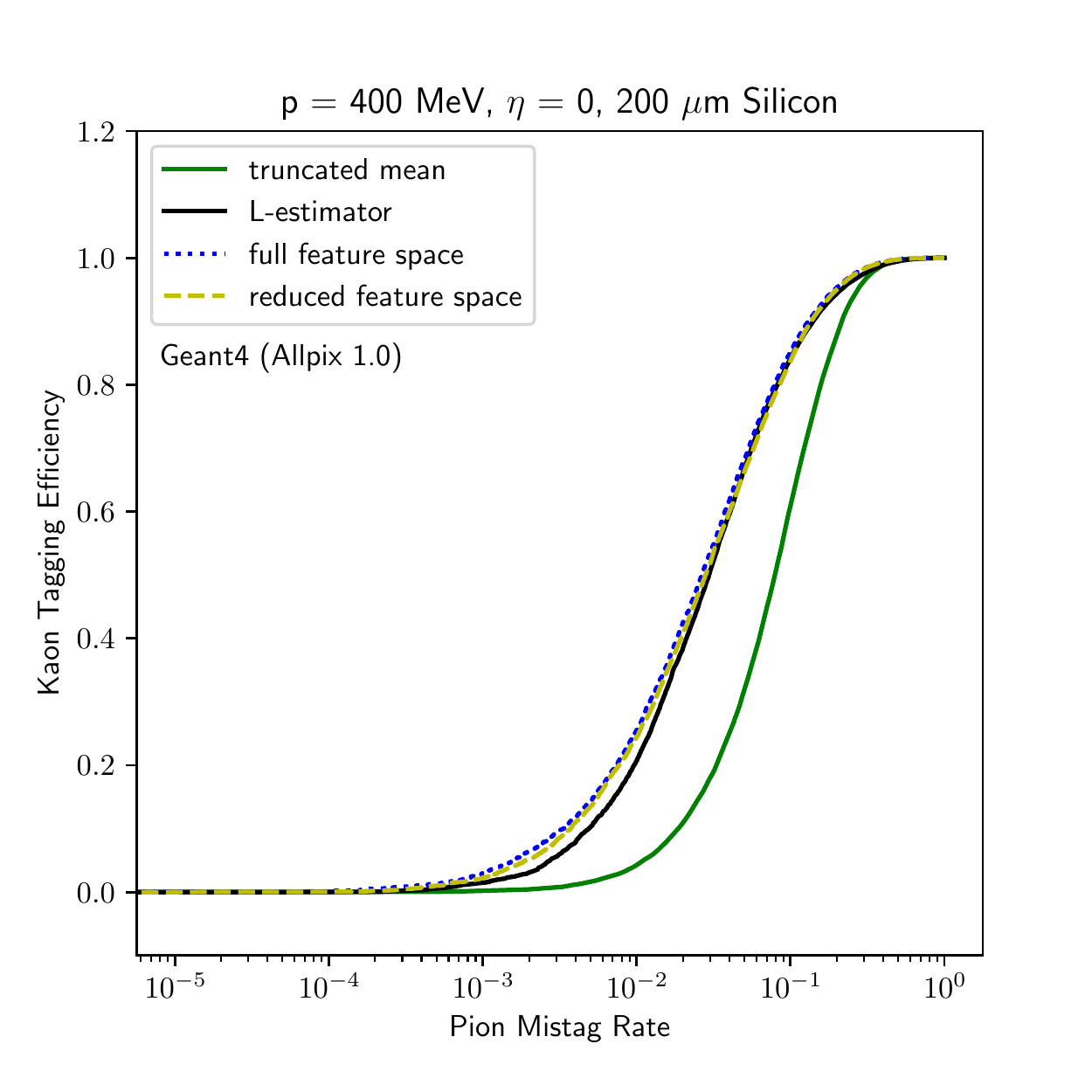}}%
		\qquad
		\subfigure{%
			\includegraphics[width=0.45\textwidth]{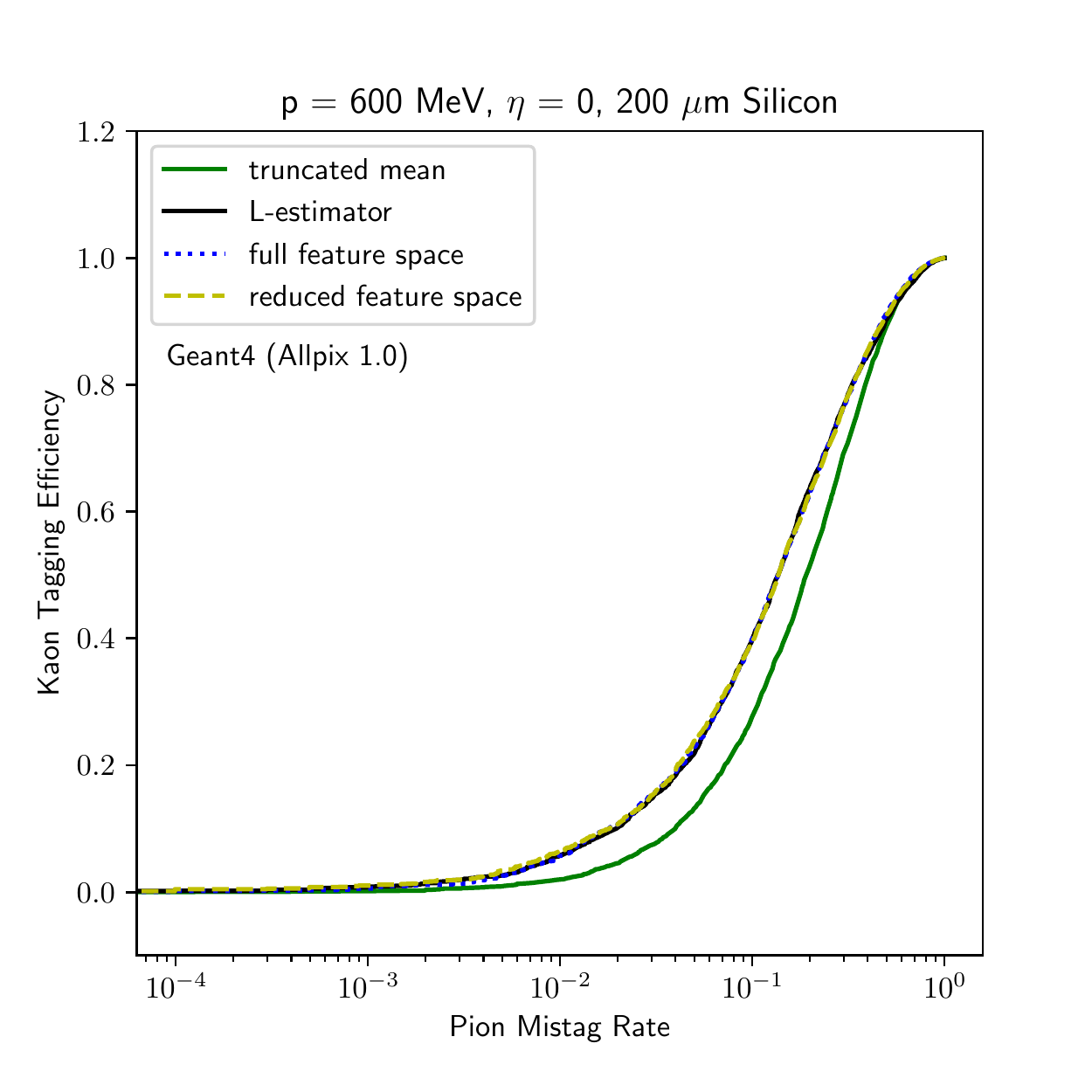}}%
	\caption{A neural network trained on the full $n$-dimensional sample points outperforms both an optimal classifier trained only with the truncated means and the optimal L-estimator at 400 MeV (left). A neural network trained on the 2-dimensional feature space consisting of $\mu_t$ and $\sigma_t$ closes the performance gap. At 600 MeV (right), all methods have the same performance.}
	\label{fig:classifier_performance}
\end{figure}

The optimal classifier for particle identification is estimated using a simple fully connected feed forward artificial neural network (NN).  Such NNs are universal function approximators~\cite{HORNIK1991251} and are empirically known to provide excellent classification performance with a limited training dataset.  A one-layer network with 100 neurons and the sigmoid activation function is trained using scikit-learn~\cite{scikit-learn}.    Figure~\ref{fig:classifier_performance} shows the performance of the NN trained on the full $n$-dimensional feature space compared with the optimal classifier using only the truncated mean\footnote{In practice, this is basically the same as a threshold on the truncated mean, but it did not have to be so if the truncated mean was not nearly monotonically related to the likelihood ratio.}.   Especially for lower momentum particles, the gap between the truncated mean classifier and the optimal classifier is large\footnote{The gap with respect to the optimal L-estimator is much smaller.}: for a pion mistag rate of 10\%, the optimal classifier has a kaon efficiency of 90\% while the truncated mean classifier has a kaon efficiency of 60\% at 400 MeV.  This performance gap can be nearly eliminated by introducing a measure of spread in addition to the truncated mean.  One possibility is to use the truncated standard deviation $\sigma_t(\vec{x})$ of each sample point. As with the truncated mean, the largest value of the sample point is removed before any statistics are calculated in order to account for the large tail of the Landau distribution.   A NN trained with only $\mu_t$ and $\sigma_t$ performs nearly the same as one trained on the full sample points as indicated in Fig.~\ref{fig:classifier_performance}.  This suggests that the reduced two-dimensional feature space contains all of the classification power of the original feature space.  In particular, truncation is a nearly lossless operation.

%, allowing us to conclude that this reduced feature contains all of the classification power of the original feature space. A performance comparison between neural networks trained on these different spaces is avaliable in Figure~\ref{fig:classifier_performance}.

%We can then define the transformation $\Phi: \mathbb{R}^n \rightarrow \mathbb{R}^2$ as follows: 
%\begin{align}
% \Phi(\vec{x}) = \begin{pmatrix}
%\mu_t(\vec{x}) & \sigma_t(\vec{x})
%\end{pmatrix}
%\end{align}

%Since artificial neural networks are universal function approximators, training one on the unprocessed sample points gives us a benchmark on optimality. That is, if we can find a minimal state representation for each track that can then be used to train a classifier with an equal performance as the neural network trained on the full sample points, then we can be sure that this simpler representation contains all of the classifying power of the full energy loss readouts.

\begin{figure}[h]%
	\centering
	\subfigure{%
		\includegraphics[width=0.45\textwidth]{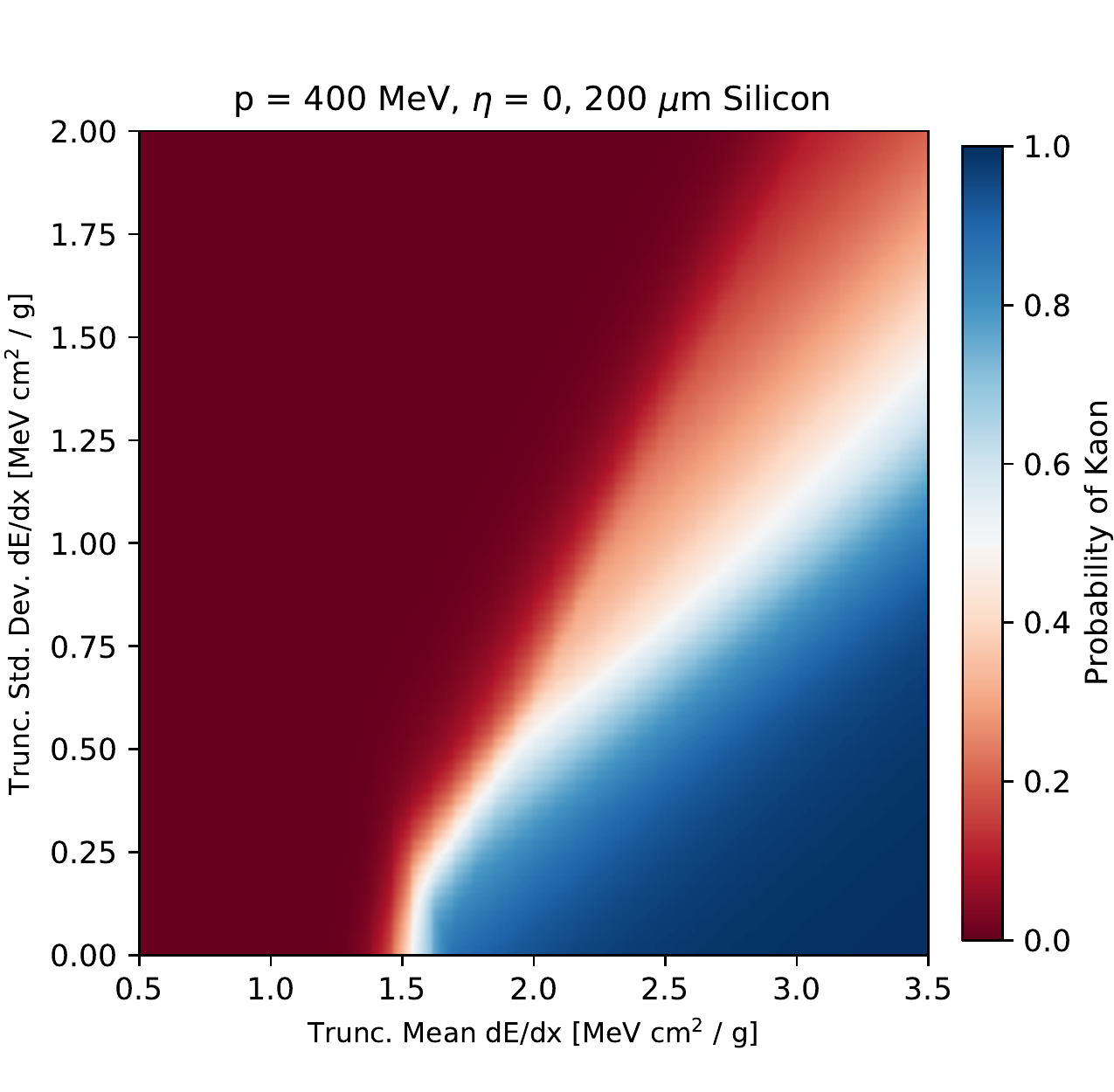}}%
	\qquad
	\subfigure{%
		\includegraphics[width=0.45\textwidth]{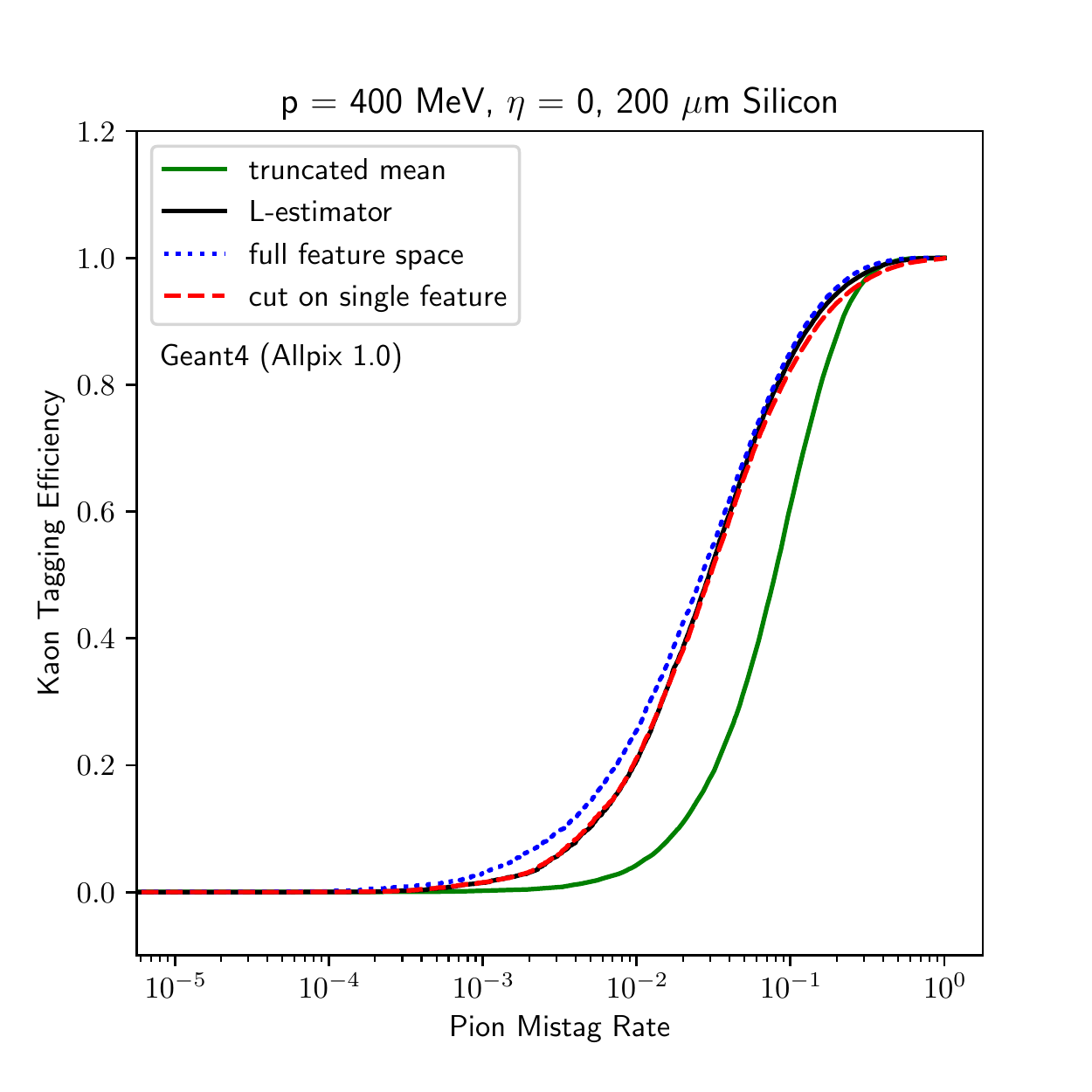}}%
	\caption{The optimal decision boundary (left) takes the form of a quadratic relationship between $\mu_t(\vec{x})$ and $\sigma_t(\vec{x})$. As a result, a linear cut on the single feature $\mu_t(\vec{x}) - \sigma_t^2(\vec{x})$ performs as well as the optimal L-estimator and nearly as well as a neural network trained on the full sample points (right).}
	\label{fig:quadratic}
\end{figure}

Reducing the original $n$-dimensional parameter space to two is a significant achievement, but the optimal classifier boundary can still be complicated.  It is therefore desirable to investigate if a simple transformation can be used to render the optimal classifier linear.  The left plot of Fig.~\ref{fig:quadratic} shows the ratio of the kaon-to-pion NN output that is trained on $\mu_t$ and $\sigma_t$.  This plot shows that sample points with high average energy loss and low spread in energy loss are the most likely to be classified as kaons, following a decision boundary roughly defined by a quadratic relationship between the truncated mean and truncated standard deviation:
\begin{align}
\label{eq:reduced}
\mu_t(\vec{x}) = \sigma_t^2(\vec{x})/\text{[MeV cm$^2$/g]} + C,
\end{align}

\noindent where $C$ is a constant.  Therefore, a simple threshold on the difference between the truncated mean and truncated variance should create nearly the same decision boundary as a NN trained on the full $n$-dimensional feature space.  The right plot of Fig.~\ref{fig:quadratic} shows a ROC curve for the full feature space compared with a simple threshold on Eq.~\ref{eq:reduced}.  The two curves are nearly identical, which shows that a one-dimensional feature is nearly sufficient to capture all of the relevant information for classification.  

%We now have the ability to map sample points of an arbitrarily large dimensionality into points in a two-dimensional feature space without any loss of optimality. However, the decision boundary formed in this reduced space could itself still be arbitrarily complicated. The next natural step would therefore be to create a feature space that is not only of low dimensionality, but also one in which the optimal decision boundary would be linear - that is, a sample point could be classified by determining on which side of a hyperplane bisecting this space it lies. Plotting the posterior probabilities outputted by a neural network trained in the current reduced feature space shows a clear decision boundary, as shown in Figure~\ref{fig:quadratic}. In this space, sample points with high average energy loss and low spread in energy loss are the most likely to be classified as kaons, following a decision boundary roughly defined by a quadratic relationship between the truncated mean and truncated standard deviation:
%\begin{align}
%\mu_t(\vec{x}) = \sigma_t^2(\vec{x}) + C
%\end{align}

%This is itself equivalent to a linear relationship between the truncated mean and truncated variance. Therefore, a linear cut along the feauture $\mu_t(\vec{x}) - \sigma_t^2(\vec{x})$ would create the same decision boundary as a neural network trained on the full dimensioned sample points (Figure~\ref{fig:quadratic}), giving us a single feature which contains the classifying power of the original feature space.

%\clearpage

Intuition for the optimality of Eq.~\ref{eq:reduced} can be derived from the ideas of sufficient statistics and exponential families~\cite{10.2307/91208}.  The optimality of truncation was already discussed in the context of $\delta$-rays in Sec.~\ref{sec:existmeth}.   While the energy loss is nearly Landau-distributed, the truncated energy loss is closer to a Gaussian distribution.   Classification with Gaussian-distributed random variables only requires the sample mean and sample standard deviation for optimal performance.  To see this, note that for a threshold $\alpha$ on the likelihood ratio, the optimal decision boundary for kaons-versus-pions in the Gaussian limit takes the following form:

\begin{align}
p_{\pi}(\vec{x}) &= \alpha p_{k}(\vec{x})\\
\frac{1}{(\sqrt {2\pi} \sigma_{\pi})^n} \exp \left(-\frac{|\vec{x} - \vec{\mu_{\pi}}|^2}{2\sigma_{\pi}^2} \right) &= \alpha \left[\frac{1}{(\sqrt {2\pi} \sigma_k)^n} \exp \left(-\frac{|\vec{x} - \vec{\mu_k}|^2}{2\sigma_k^2} \right)\right],
\end{align}

\noindent where $\vec{\mu}_k,\sigma_k$ are respectively the mean and standard deviation for kaons and analogously $\vec{\mu}_\pi,\sigma_\pi$ are for the defining parameters for pions. Taking the logarithm of both sides and simplifying yields the following:

%The optimality of this reduced feature space can be derived from the pion and kaon energy loss distributions directly. As the degree of truncation increases, the Landau energy loss distributions begin to converge to become rougly Gaussian (Figure~\ref{fig:gaussian}). It is then reasonable to approximate both the pion and kaon energy loss distributions as $n$-dimensional multivariate Gaussians with independent and identically distributed features (with parameters $\mu_\pi,\sigma_\pi$ and $\mu_k,\sigma_k$ respectively). Since the first and second moments are sufficient statistics for the normal exponential family, one can optimally predict the class of a given $n$-dimensional sample point even after it undergoes the feature map $\Phi: \mathbb{R}^n \rightarrow \mathbb{R}^2$ described above.

\begin{align}
\label{eq:simplified}
\left(\frac{\mu_{\pi}}{\sigma_{\pi}^2} - \frac{\mu_{k}}{\sigma_{k}^2}\right)\mu_1(\vec{x}) - 
\frac{1}{2}\left(\frac{1}{\sigma_{\pi}^2} - \frac{1}{\sigma_{k}^2}\right)\mu_2(\vec{x}) 
= \frac{1}{2}\left(\frac{\mu_{\pi}^2}{\sigma_{\pi}^2} - \frac{\mu_{k}^2}{\sigma_{k}^2}\right) + \log\frac{\alpha\sigma_\pi}{\sigma_k},
\end{align}

\noindent where $\mu_1(\vec{x})$ and $\mu_2(\vec{x})$ denote the first and second raw moments of $\vec{x}$, respectively. Equation~\ref{eq:simplified} corresponds to a decision boundary of the form $\mu_1(\vec{x}) + w \mu_2(\vec{x}) = b$ for some weight $w$ and bias $b$ determined by the distributions themselves.  This resembles the form of Eq.~\ref{eq:reduced}. The optimal value of the weight $w$ can further be tuned as another hyper-parameter. For the problem of pion vs kaon classification, we found the optimal value to be near 1.2 g / MeV cm$^2$. The change from this weight to the weight of 1 g / MeV cm$^2$ used in our approximation causes a negligible decrease in classification power, corresponding to a drop of less than 0.01 in the area under the corresponding ROC curve.  Furthermore, the functional form in Eq.~\ref{eq:reduced} continues to be close to optimal even when there are more than four layers, though there is some degradation when there are many more than four layers.  

\begin{figure}[h]%
	\centering
	\subfigure{%
		\includegraphics[width=0.45\textwidth]{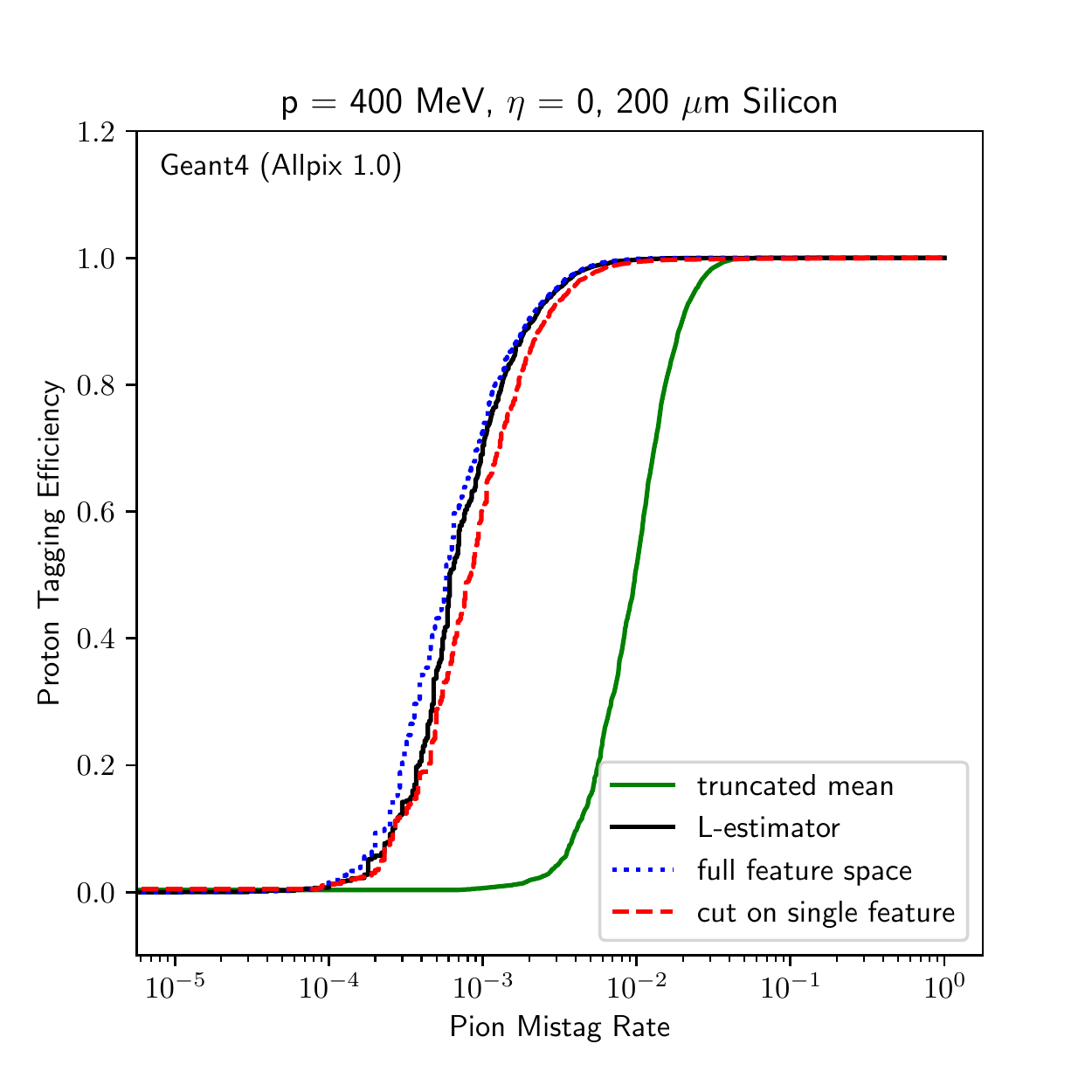}}%
	\qquad
	\subfigure{%
		\includegraphics[width=0.45\textwidth]{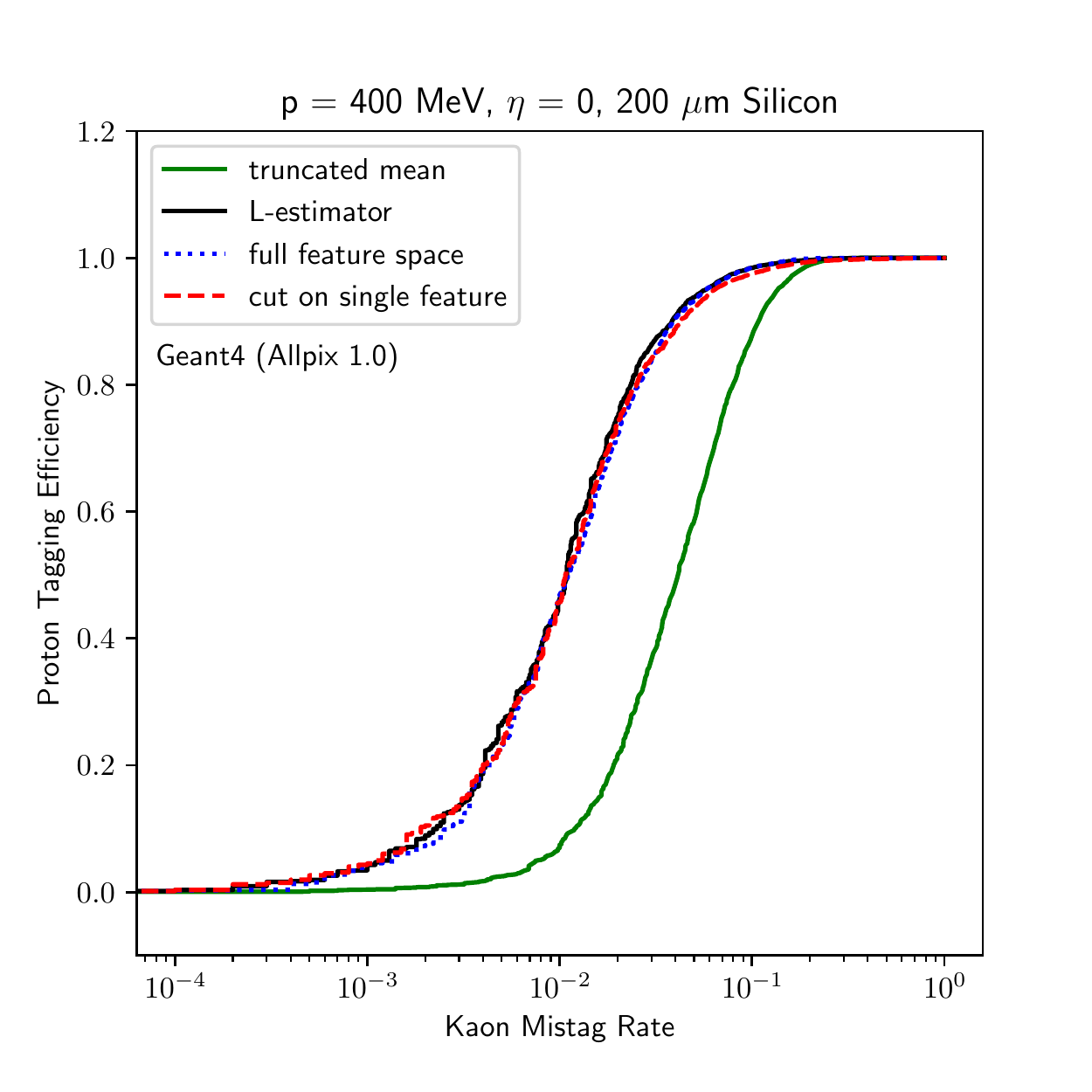}}%
	\caption{Both a linear cut on the single feature $\mu_t(\vec{x}) - \sigma_t^2(\vec{x})$ and the optimal L-estimator come close to matching the performance of a neural network trained on the full sample points for the tasks of distinguishing protons from pions (left). All three methods have the same performance for the task of distinguishing protons from kaons (right).}
	\label{fig:proton}
\end{figure}

This technique can be extended to identification of tracks from particles besides pions and kaons. We observe similar performance when our model is used to distinguish protons from either pions or kaons (Figure~\ref{fig:proton}). 

%We can therefore conclude that, with adequate truncation, the optimal decision boundary can be very closely approximated with a linear cut in this reduced feature space.

%This approximation begins to diverge fron the optimal classifer, however, as the number of detector layers grows. Still, for detectors such as the ATLAS pixel detectors that have setups similar to what we used here, the optimal classifier is well-approximated using this method.

%We also observe similar performance when the number of layers of the detector is increased (Figure~\ref{fig:double}), confiming that our feature map can reduce abitrarily large sample points.

%\begin{figure}[h]%
	%\centering
	%\subfigure{%
		%\includegraphics[height=2in]{../figures/400/nn_tm_cut_double.png}}%
	%\caption{Doubling the dimensionality of the sample points from four to eight does not signifiantly impact classsifer performance.}
	%\label{fig:double}
%\end{figure}

%-------------------------------------------------------------------------------
\section{Conclusions}
\label{sec:concl}

The energy deposited by charged particles in silicon tracking detectors is useful for identifying the particle type when $\beta\gamma<1$.  The truncated mean has been used by many analyses at the LHC and previous experiments.  We have shown that truncation is a nearly lossless operation on the set of dE/dx values obtained from multiple tracking layers.  However, the truncated mean alone is not sufficient to fully capture all of the available information.  By adding a measure of spread as a second feature, we are able to achieve the optimal classification.  A simple difference of the truncated mean and variance is nearly sufficient to fully capture all of the available information.  Similar results are observed when varying the particle momentum as well as the particle types.  Hopefully the techniques developed here will be useful for fully exploiting the data at the LHC for both measurements of fragmentation and other quantum chromodynamical processes as well as searches for new particles beyond the Standard Model that are charged and slow moving.
%-------------------------------------------------------------------------------

%\clearpage

\section{Acknowledgments}

We would like to thank Maurice Garcia-Sciveres for comments on the manuscript.  This work was supported by the U.S.~Department of Energy, Office of Science under contract DE-AC02-05CH11231.

\clearpage

\appendix

%\section{Additional Plots}

%\begin{figure}[h]%
	%\centering
	%\subfigure{%
		%\includegraphics[height=2in]{../figures/400/trunc_0.png}}%
%	\qquad
%	\subfigure{%
	%	\includegraphics[height=2in]{../figures/400/trunc_1.png}}\\
%	\subfigure{%
	%	\includegraphics[height=2in]{../figures/400/trunc_3.png}}%
%	\qquad
%	\subfigure{%
	%	\includegraphics[height=2in]{../figures/400/trunc_5.png}}%
%	\caption{As the number of truncated terms increases, the pion and kaon energy loss distributions begin to become approximately Gaussian, as shown on eight-dimensional sample points with no truncation (top left), one truncation (top right), three truncations (bottom left), and five truncations (bottom right).}
%	\label{fig:gaussian}
%\end{figure}

%\bibliographystyle{ieeetr}
\bibliographystyle{elsarticle-num}
\bibliography{myrefs.bib}{}

\end{document}